\documentclass[12pt]{iopart}

\eqnobysec
\usepackage[]{graphicx}
\usepackage{amssymb}
\begin{document}
\title[Semi-relativistic wave-phase approximation ]{\begin{center}\Large Semi-relativistic wave-phase approximation for two-body spinless bound states  in 1+1 dimensions \end{center}}
\author{K-E Thylwe and S Belov}
\address {{\small  KTH-Mechanics, Royal institute of Technology,}
 {\small S-100 44 Stockholm, Sweden.} }
\vspace{0cm}

\begin{center}
$\mbox{Working copy:\;{\today}} $ 
\vspace{1cm}

\noindent {\large \bf Abstract} 
\end{center}
An approximate quantum-mechanical two-body equation for spinless particles incorporating relativistic kinematics is derived. The derivation is based on the  relativistic energy-momentum relation $mc^{2}+\epsilon = \sqrt{m^{2}c^{4}+p^{2}c^{2}}+V$ for each single particle, where $mc^2$ is the particle rest mass energy, $p$ its linear momentum, $\epsilon$ its dynamical energy, and $V$  being the time-like vector interaction potential. The resulting two-body equation assumes rapid wave oscillations in a  single, slowly varying  potential well. A Bohr-Sommerfeld-type quantization condition is obtained. The approximation is compared to exact results for the harmonic potential.
\vspace{0.5cm}
\section{lntroduction} \label{S1}
Approaches to derive  two-body equations valid in the limit of relativistic quantum mechanics may start from the field-theoretic Bethe-Salpeter equation \cite{Lucha}. By a series of approximations involving the neglect of spin effects one then ends up with Hamiltonian equations based on the classical energy-momentum relation $M=\sqrt{m^{2}c^{4}+p^{2}c^{2}}+V$ \cite{Ikhdair,Rajabi}, where $M=\epsilon + mc^{2}$, and $\epsilon$ is the non-relativistic energy. In an  approach based on classical relativity, the square root in such Hamiltonians become a quantal operator causing intricate problems, not only for the difficulty of explaining spin effects of the particles.  The approach by Dirac, resulting in the celebrated Dirac  equation is one fruitful result for spin-$\frac{1}{2}$ particles of  such considerations. However, the Dirac equation originally describes just a single particle in a potential. The present approach ignores spin effects.

The present study is based an amplitude-phase decomposition of the wave solution \cite{Milne}, and  generalizes the spinless Schr\"{o}dinger equation.  Waves of the form $A(x)\exp(\pm\rmi S(x)/\hbar)$ are assumed, where $\hbar$ is Planck's constant. A relation between the amplitude $A$ and the phase derivative $S'(x)/\hbar$ is obtained, resulting in a single differential equation for the amplitude $A$.  By including leading-order relativistic contribution of the phase factor $\exp(\rmi S(x)/\hbar)$ a quantization condition similar to a Bohr-Sommerfeld quantization condition is derived \cite{T85,KL81,T15a,Friedrich04}.

The resulting leading-order relativistic wave-phase contribution approximates the two-body operators $\sqrt{m_{j}^{2}c^{4}+\hat{p}^{2}c^{2}}$ with masses $m_{1}$ and $m_{2}$.
 The method is presented in 1+1 dimensions with a numerical application to the harmonic potential. Note that such a potential has not any discrete spectrum in the Dirac and Klein-Gordon theories, but they should still exist \cite{Lucha}.  
 
The  suggested local approximation of the same problem given  in \cite{Ikhdair} is frequently applied by several authors \cite{Rajabi}. Comments on this approach together with rigorous bound-state energy estimates for radial potentials  are presented in a series of articles by Lucha and Sch\"{o}berl \cite{Lucha}. As pointed out by this reference, a disadvantage of the approach in \cite{Ikhdair}, and its  applications  \cite{Rajabi}, is its inapplicability to classically confining systems, where the potential increases indefinitely towards infinite particle separations. The present method does not suffer from this deficiency.

Section 2 presents the basic two-body equations in 1+1 dimensions and the relevant expansion approximation of the square-root operators. The fundamental  amplitude-phase solutions are introduced in the non-relativistic limit. Section 3 introduces the dominating relativistic corrections in two steps; firstly the main operator action on the phase, secondly the the phase corrections  are transferred to the amplitude function by imposing a phase-amplitude relation. Numerical details and illustrations are discussed in section 4. Section 5 summarizes the main conclusions.

\section{Quantum-mechanical two-particle equations with relativistic corrections} 
For two particles with masses $m_{1}$ and $m_{2}$ the relative motion along an $x$-axis in a center-of-mass reference system  is
such that
\begin{equation}
x_{1}=\frac{m_{2}}{m} x,\;\;x_{2}=-\frac{m_{1}}{m}x,\;\;m=m_{1}+m_{2},
\end{equation}
where $x=x_{1}-x_{2}$ defines the relative position of the particles, and $m$ is the total rest mass. As in classical mechanics, the linear momenta relative to the centre of mass are related by the vector equation ${\bf p}_{2}=-{\bf p}_{1}$, and one can define a single operator symbol $\hat{p}=\hat{p}_{1}$. The wave function $\psi$ satisfies the equation
\begin{equation}
\left(\sqrt{m_{1}^{2}c^{4}+\hat{p}^{2}c^{2}}+\sqrt{m_{2}^{2}c^{4}+\hat{p}^{2}c^{2}}+V(\hat{x})\right)\psi=M \psi,
\label{Meq2}
\end{equation}
where $M=mc^{2}+\epsilon$ is the total energy.

The Schr\"{o}dinger correspondence rule specifies equation (\ref{Meq2}) in treating $\hat{p}$ as a differential operator $\hat{p}^{2}=-\hbar^{2}\rmd^{2}/\rmd x^{2}$ and keeping $\hat{x}=x$ as a variable coordinate. Equivalently, in a conjugate approach $\hat{x}$ is an operator $\hat{x}^{2} = -\hbar^{2}\rmd^{2}/\rmd p^{2}$, while $\hat{p}=p$ is treated as a variable. Since the harmonic-oscillator potential case  is simply and exactly solvable by standard methods in the momentum representation, its results will be used here as a reference case.

The operator
\begin{equation}
\hat{\Sigma}=\left(\sqrt{m_{1}^{2}c^{4}+\hat{p}^{2}c^{2}}+\sqrt{m_{2}^{2}c^{4}+\hat{p}^{2}c^{2}}\right),
\label{Oexp}
\end{equation}
with which  (\ref{Meq2}) can be written
\begin{equation}
\hat{\Sigma} \psi =(mc^{2} + \epsilon-V) \psi, 
\label{Deq}
\end{equation}
is considered as defined by the formal expansion
\begin{equation}
\hat{\Sigma}= mc^{2}+ \mu c^{2}\sum_{j=1}^{\infty} {\frac{1}{2}\choose j}\frac{\eta_{2j}}{c^{2j}\mu^{2j}} \hat{p}^{2j} =
mc^{2}+ \frac{\hat{p}^{2}}{2\mu} - \frac{\eta_{4}}{8\mu^{3}c^{2}}\hat{p}^{4} + \cdots,
\label{Oapp}
\end{equation}
with the reduced ($\mu$) and total ($m$) masses
\begin{equation}
\mu=\frac{m_{1}m_{2}}{m},\;\; m=m_{1}+m_{2},
\end{equation}
defined, together with the mass distribution coefficients
\begin{equation}
\eta_{2j}=\left(\frac{m_{1}}{m}\right)^{2j-1}+\left(\frac{m_{2}}{m}\right)^{2j-1}.
\end{equation}
In particular, for the leading two terms (with $j=1,2$) one has
\begin{equation}
\eta_{2}= 1, \;\; \eta_{4}= \left(\frac{m_{1}}{m}\right)^3+\left(\frac{m_{2}}{m}\right)^{3}.
\end{equation}

With the expansion (\ref{Oapp})  inserted into (\ref{Deq}), one obtains the  Schr\"{o}dinger equation, i.e.
\begin{equation}
- \frac{\rmd^{2}}{\rmd x^{2}}\psi_{0}=\frac{2\mu}{\hbar^{2}}\left(\epsilon-V\right) \psi_{0}. 
\label{psiS}
\end{equation}

For the Schr\"{o}dinger equation it is well known that the amplitude-phase method can be applied in a rigorous way, without any approximations. This is illustrated as follows:

 An amplitude-phase assumption of fundamental solutions taking the form
\begin{equation}
\psi_{0}=A\exp{(\pm\rmi S/\hbar)}
\label{APbase}
\end{equation}
is inserted into  (\ref{psiS}), yielding
\begin{equation}
\fl \left(-A''+\left( \frac{S'}{\hbar}\right)^{2}A\mp\rmi \left(2A'\frac{S'}{\hbar}+A\frac{S''}{\hbar}\right) \right)\exp{(\pm\rmi S/\hbar)}=\frac{2\mu}{\hbar^{2}}\left(\epsilon-V\right)A\exp{(\pm\rmi S/\hbar)}.
\label{APS}
\end{equation}
As is typical in the amplitude-phase approach, a relation between amplitude and phase functions is defined. Here, it is convenient to eliminate the imaginary terms in (\ref{APS}) by the condition
\begin{equation}
(2A'S'+AS'')\hbar^{-1}= A^{-1}\left(A^{2}S' \hbar^{-1}\right)'=0.
\end{equation}
Hence, a replacement 
\begin{equation}
\frac{S'}{\hbar} \rightarrow A^{-2}
\label{APrel}
\end{equation}
can be made in (\ref{APS}).
Note that an unimportant multiplying factor in (\ref{APrel}) is here put to unity. The Schr\"{o}dinger equation is then (exactly) transformed into the non-linear amplitude equation:
\begin{equation}
A''+\frac{2\mu}{\hbar^{2}}\left(\epsilon-V\right)A=A^{-3}.  \;\;\mbox{(S)}
\label{APSeq}
\end{equation}
In the nonrelativistic and spinless two-body problem, the amplitude-phase
 equation (\ref{APSeq}) is exact. Physically, one sees no effects from different combinations of masses representing the same reduced mass $\mu$. 
 The numerical computation of (\ref{APSeq}) is typically performed with initial conditions in the oscillating (well) region such that $A'\approx A'' \approx 0$. It is assumed here that only one such oscillating region exists. In non-relativistic problems with slowly varying potentials $V$ the amplitude function is then smooth and slowly varying inside the well, but then turns exponentially increasing as the wave function penetrates into a 'classically forbidden' region, where $V>\epsilon$. 
 
 It is in these transitions from 'allowed' to 'forbidden' classical regions that the term $A''$ becomes large and most important. Note that as $A$ increases, the phase of the wave (represented by the integral $S/\hbar =\int^{x} A^{-2}\rmd x$) converges faster than $A$ increases. This guarantees that the fundamental solutions (\ref{APbase}) are still useful for representing a decaying wave function outside of the potential well by a suitable linear combination. This idea will next be used to include relativistic contributions in an approximate way.

\subsection{Relativistic wave-phase approximation}
When including relativistic terms it is argued here that an important class of physically potentials are smooth and slowly varying over the relevant region where the wave is oscillating. The derivative of the potential is then small in some sense in the relevant well region. Another argument is related to  semiclassical approximations \cite{Friedrich04}; the oscillations are assumed being fast in some sense. The smallness of $\hbar$ then becomes important, and this focuses on the oscillating factors $\exp(\pm \rmi S/\hbar)$ in a wave representation like (\ref{APbase}). 

The considerations mentioned here become important in the  process of identifying expansion terms in (\ref{Oexp}) to be included in the amplitude equation generalizing (\ref{APSeq}). 
In this selection procedure  one considers the wave primarily in the interior of the  region of oscillation. As the (non-relativistic) semiclassical Bohr-Sommerfeld quantization condition indicates, the amplitude of the wave plays a minor role. The non-oscillating regions of the wave (where solutions stop oscillating) are less sensitive for calculations bound-state spectra. 

The most important aspect of such a 'wave-phase' approximation is that the fundamental solutions become well behaved  also in the non-oscillating regions (to represent there the decaying wave) without further matching techniques. The amplitude-phase method for the Schr\"{o}dinger equations is defined (exactly) in the oscillating region and automatically transfers this wave solution into the 'exponential' region in a useful form. This behavior is controlled by the amplitude, which is (like the potential) slowly varying inside the potential well, but as the wave approaches the transition points to the 'forbidden' regions it makes the phase slowing down by allowing the amplitude to grow (for single-well states) indefinitely.

In the rapidly oscillating region the leading contributions of the operator  (\ref{Oexp}) in this approximation are 
\begin{eqnarray}
\hat{\Sigma} A\rme^{\pm \rmi S/\hbar} \approx &&\left(\sqrt{m_{1}^{2}c^{4}+S'^{2}c^{2}}+\sqrt{m_{2}^{2}c^{4}+S'^{2}c^{2}} \right)A\rme^{\pm \rmi S/\hbar} \nonumber \\&&- \frac{\hbar^{2}}{2\mu}\left[ A'' \mp \rmi \left(2A'\frac{S'}{\hbar}+A\frac{S''}{\hbar}\right) \right]\rme^{\pm\rmi S/\hbar}.
\label{ASexp}
\end{eqnarray}
In (\ref{ASexp}) all the important terms in the Schr\"{o}dinger operator in (\ref{APS}) are included. Terms including derivatives of the amplitude function $A$ were not successfully summed up in this study and the approximation consists of just the non-relativistic differential amplitude terms. However, as argued earlier, derivatives of the amplitude function are assumed to play a minor role in the interior of the potential well and the wave oscillation. The main relativistic corrections retained relate to the oscillatory phase factors $\exp(\pm \rmi S/\hbar)$. This contribution is easily summable through the formal expansion (\ref{Oapp}) operating on  $\exp(\pm \rmi S/\hbar)$ alone.

From the second term in (\ref{ASexp}) one concludes that the relation between amplitude and phase (\ref{APrel}) can again be implemented, which in turn means that the amplitude also include relativistic corrections. The resulting approximate differential equation is given by
\begin{equation}
\fl A'' +\frac{2\mu}{\hbar^{2}} \left[\epsilon-V +mc^{2}-\left(\sqrt{m_{1}^{2}c^{4}+A^{-4}\hbar^{2}c^{2}}+\sqrt{m_{2}^{2}c^{4}+A^{-4}\hbar^{2}c^{2}} \right)\right]A=0.\;\;  \;\;\mbox{(WP)}
\label{APeq}
\end{equation}
In the non-relativistic limit $c \rightarrow \infty$  equation (\ref{APeq}) reduces to (\ref{APSeq}).

To understand the assumed slow variations of the potential one should relate the range of the potential $r_{*}$, say, to that of the relevant Compton wave length $\hbar/(\mu c)$. The potential range is assumed significantly larger that the Compton length for the relative system. With a rescaling of the space variable by the substitution $z=x/x_{*}$, where $x_{*}$ is arbitrary, equation (\ref{APeq}) becomes
\begin{equation}
\fl \frac{\rmd^{2}A}{\rmd z^{2}} +\frac{2\mu x_{*}^{2}}{\hbar^{2}} \left[\epsilon-V +mc^{2}-\left(\sqrt{m_{1}^{2}c^{4}+A^{-4}\hbar^{2}c^{2}}+\sqrt{m_{2}^{2}c^{4}+A^{-4}\hbar^{2}c^{2}} \right)\right]A=0.
\label{APeqz}
\end{equation}

\section{Calculations and illustrations}

The numerical solution of (\ref{APeq}) uses a convenient linear combination of fundamental amplitude-phase solutions (\ref{APbase})  for single-well potentials \cite{T15a}
\begin{equation}
\psi = A \sin \left(\int_{-\infty}^{x}A^{-2}\rmd x\right),\;\; \psi(\pm\infty)=0.
\label{wavedef}
\end{equation}

Exact initial values of $A$ and $A'$ in solving the amplitude equation (\ref{APSeq}) in the non-relativistic case are not needed \cite{T15a}. However, since the amplitude differential equation is nonlinear, its solutions may be qualitatively different for different initial conditions. Most frequently it is preferable to have the amplitude slowly varying in the 'classically allowed' region of the potential. The wave representation is however still the same, independent of initial conditions.

The new amplitude equation (\ref{APeq}) is an approximation of the relativistic (spinless) two-body equation (\ref{Meq2}). The exact independence of initial conditions is lost. This means that the computational advantage of slowly varying amplitude function in the non-relativistic case turns more essential in the relativistic case. since the derivation requires this amplitude behavior in the oscillating ragion.

Assuming $A$ having small derivatives inside the classically allowed region, equation (\ref{APeq}) can be approximated by
\begin{equation}
 \left[mc^{2}+\epsilon-V -\left(\sqrt{m_{1}^{2}c^{4}+A^{-4}\hbar^{2}c^{2}}+\sqrt{m_{2}^{2}c^{4}+A^{-4}\hbar^{2}c^{2}} \right)\right] \approx 0.
\label{adAPeq}
\end{equation}
Using this approximation, and also assuming the integration being initiated at the (single) minimum of the potential, here chosen to be at the origin $x=0$, one finds an initial value of $A=A_{0}$ by solving the equation:
\begin{equation}
\sqrt{m_{1}^{2}c^{4}+A_{0}^{-4}\hbar^{2}c^{2}}+\sqrt{m_{2}^{2}c^{4}+A_{0}^{-4}\hbar^{2}c^{2}}= mc^{2}+\epsilon-V_{0},
\label{adAPeqsol}
\end{equation}
where the right hand member is constant for a given guess of the energy $\epsilon$.
Choosing $m_{1}\geq m_{2}$ the algebraic equation has a solution
\begin{equation}
\frac{\hbar^{2}c^{2}}{A_{0}^{4}}=\frac{1}{4}\left[\frac{\left(mc^{2}+\epsilon-V_{0}\right)^{2}-m_{1}^{2}c^{4}+m_{2}^{2}c^{4}}{mc^{2}+\epsilon-V_{0}}\right]^{2}-m_{2}^{2}c^{4},
\label{Ainit}
\end{equation}
from which $A_{0}$ is determined.
Initial conditions for the amplitude function becomes $A=A_{0}$, from (\ref{Ainit}), and $A'=0$. The integration runs in two directions from the potential minimum at $x=0$. In fact, for symmetric potentials one needs  integration only in one direction.

Together with this integration of the differential equation (\ref{APeq}), an integration of the phase integrand $A^{-2}$ is practical. The phase integrand tends rapidly to zero as one enters into a classically forbidden region and indicates where the the integration terminates. The result of this integration in a symmetric case is half the phase integral across the well. Hence, the total phase integral in (\ref{wavedef}) for bound states satisfies:
\begin{equation}
\int_{-\infty}^{+\infty}A^{-2}\rmd x=\left(n+1 \right)\hbar,\;\;n=0, 1, \cdots.
\label{BS}
\end{equation}
The quantum number $n$ can be interpreted as the number of nodes of the wave.

\begin{table}
\caption{\small Bound state energies for $V=\beta x^{2}/2$ obtained by an exact relativistic model ($\epsilon_{R}$) is compared with those of the present wave-phase (WP) approximation ($\epsilon_{QP}$) from equation (\ref{BS}) and the exact non-relativistic model ($\epsilon_{NR}$). The potential parameter is fixed to $\beta=1$ and mass parameters  $\mu$ and $m_{1}$ are varied. Nodal quantum numbers $'n'$ are added.}
\small
\begin{center}
\begin{tabular}[t]{ccclll} \hline
\hline
$n$&$\mu,\ m_{1}$ & $\epsilon_{\rm R}$& &  $\epsilon_{WP}$ & $\epsilon_{\rm NR}$  \\
\hline
\hline
&&&&&\\
0&5, 10& 0.222686 &&0.2226 &  0.224 \\
10&5, 10&  4.509109 &&4.506 & 4.7 \\
20&5, 10&  8.515499 && 8.512 & 9.2 \\ 
&&&&&\\
0&5, 100& 0.220584 && 0.2202 &  0.224 \\
10&5, 100&  4.183784 && 4.174 &  4.7 \\
20&5, 100& 7.591316 && 7.580 &  9.2 \\ 
&&&&&\\
0&1, 2& 0.480097 && 0.476 &  0.50 \\
10&1, 2&  7.965927 && 7.914 & 10.5 \\
20&1, 2&  13.646201 && 13.59 & 20.5 \\ 
&&&&&\\
0&1, 10&  0.455021 &&0.442 &  0.50 \\
10&1, 10&  6.846005 && 6.749 &  10.5 \\
20&1, 10& 11.760647 && 11.66 &  20.5 \\ 
\hline
\end{tabular}
\vspace{5mm}
\end{center}
\label{table1}
\end{table}

The table shows exact bound state energies compared with those calculated from the present wave-phase (WP) approximation and the non-relativistic (NR) (Schr\"{o}dinger) approximation.  The harmonic oscillator potential is the same in all calculations, with $\beta=1$.

The first two groups of entries refer to a system of  a reduced mass $\mu=5$, but with two different mass combinations. The non-relativistic theory does not distinguish between these two mass cases. Similarly, the last two groups of entries refer to systems with reduced mass $\mu=1$ and two other mass combinations. 

A first general observation of semi-relativistic exact calculations is that equal-mass systems have spectra closer to non-relativistic predictions than light-heavy mass systems. 

A second general observation is that the WP approximation is better for larger reduced masses.

A third general observation is that relativistic effects become more significant at higher nodal quantum numbers $n$.
The non-relativistic approximation $\epsilon_{NR}$ is surprisingly accurate for the ground states $n=0$ of equal-mass systems, and also for higher states with $\mu=5$.
 
As for the present WP approximation, its predictions are consistently lower than the exact predictions, by a factor in the range of 95-100\%. WP predictions are more accurate for systems with larger reduced masses. Still, the WP approximation is predictive for several tens of  excited bound states.

\section{Conclusions}
 The present relativistic wave-phase approximation is based on the amplitude-phase method adjusted to (and approximating) the spinless two-body Hamiltonian by assuming $\hbar$ being small and the range of the well being large compared to the Compton wave length.  Relevant two-body features of the harmonic-oscillator bound-state spectrum seem to be well predicted by this approximation. The WP approximation seem to be applicable to sub-atomic and nuclear spectra whenever spin-$0$ states are considered.
 
 The method relies on specific initial conditions of its basic non-linear equation. Such conditions are presented here for non-singular potential wells only.

\end{document}